\documentclass{article}
\usepackage[utf8]{inputenc}
\usepackage{amsthm}
\usepackage{graphicx,ae,aecompl}
\usepackage{amssymb}
\usepackage{amsmath}
\usepackage{subfigure}
\usepackage{algorithmic}
\usepackage[boxruled,linesnumbered]{algorithm2e}
\usepackage{makeidx}
\usepackage{float}
\usepackage{scalefnt}
\usepackage{indentfirst}

\usepackage[left=2.5cm,top=2.5cm,right=2.5cm,bottom=2.5cm]{geometry}

\title{Simulation studies to compare bayesian wavelet shrinkage methods in aggregated functional data}
\author{Alex Rodrigo dos S. Sousa \\ State University of Campinas, Brazil \\ }
\date{}

\begin{document}

\numberwithin{equation}{section}
\numberwithin{table}{section}
\numberwithin{figure}{section}

 \maketitle
    \begin{abstract}
     The present work describes simulation studies to compare the performances of bayesian wavelet shrinkage methods in estimating component curves from aggregated functional data. To do so, five methods were considered: the bayesian shrinkage rule under logistic prior by Sousa (2020), bayesian shrinkage rule under beta prior by Sousa et al. (2020), Large Posterior Mode method by Cutillo et al. (2008), Amplitude-scale invariant Bayes Estimator by Figueiredo and Nowak (2001) and Bayesian Adaptive Multiresolution Smoother by Vidakovic and Ruggeri (2001). Further, the so called Donoho-Johnstone test functions, Logit and SpaHet functions were considered as component functions. It was observed that the signal to noise ratio of the data had impact on the performances of the methods. 
        
    \end{abstract}

\section{Introduction}
The statistical problem of estimating component curves from aggregated curves, also known as calibration problem, has been widely studied in recent years in several areas of science. In Chemometrics, for example, there is an interest in estimating absorbance curves of the constituents of a substance from samples of absorbance curves of the substance itself. In this case, the substance's absorbance curve is formed by the linear combination of the absorbance curves of its constituents, according to the Beer-Lambert Law (Brereton, 2003). Another example appears in the study of electricity consumption in a given region, in which the energy consumption curve over a period of time is composed of the aggregations of the individual consumption curves of households and establishments (Dias et al., 2013).

The first proposed methods to estimate component curves from aggregated data approach this problem in a multivariate way, since, in practice, we observe such curves in a finite number of locations. Thus, it is possible to see the observations as a random vector with a certain correlation structure between nearby locations. Methods based on Principal Components Regression (PCR) by Cowe and McNicol (1985) and Partial Least Squares Regression (PLS) by Wold et al. (1983) have been successfully proposed in applications. Bayesian multivariate approaches was also proposed by Brown et al. (1998a,b) and Brown et al. (2001).

Methods based on functional data analysis to the calibration problem were proposed later by Dias et al. (2009) and Dias et al. (2013), taking into account the functional structure of the observations. In this way, each component curve can be represented in terms of some convenient function basis, such as splines basis for example, so that the problem of estimating the curve becomes a finite-dimensional problem of estimating the coefficients of the basis expansion. In this functional approach, one possibility is to expand the functions in wavelet basis, recently proposed by Sousa (2022). An advantage of such an approach is the well localization property of the wavelet coefficients and the possibility of estimating important characteristics of a function, such as peaks, oscillations, discontinuities, among others, through wavelet coefficients. Further, the sparsity of the wavelet coefficients allow to identify these function features by the magnitude of the nonzero coefficients at their localizations. See Vidakovic (1999) for details and properties of wavelets and its application on statistical modelling.

Sousa (2022) proposes the expansion of the component curves by wavelet basis and, to estimate the wavelet coefficients, the use of a bayesian approach by considering, for each wavelet coefficient, a prior distribution composed of a point mass function at zero and the logistic distribution. Thus, the Bayesian rule for estimating coefficients under quadratic loss function assumption is the  posterior expected value of the wavelet coefficient. In fact, the associated rule acts by shrinking empirical wavelet coefficients and reducing the present noise effects on the coefficients. This kind of estimator is also called shrinkage rules, see Donoho and Johnstone (1994a,b and 1995) and Donoho (1993a,b and 1995a,b)  for details about wavelet shrinkage procedures. Further, in the work by Sousa (2022), a simulation study was carried out comparing the proposed method with the expansion of the component curves by B-splines basis. It was concluded that the proposed method performed better in terms of mean squared error than the method based on B-splines  for curves with characteristics such as discontinuities, peaks and oscillations. 

Although the study by Sousa (2022) indicates the feasibility of applying the wavelet basis expansion and the use of the Bayesian shrinkage rule under logistic prior to estimate the wavelet coefficients of component curves from aggregated curves, it would be interesting a study that compares different wavelet shrinkage methods in estimating component curves. In this sense, this work intends to compare bayesian methods of wavelet shrinkage in the problem of estimating component curves from aggregated curves. For that, five methods were considered, the rule based on the logistic prior of Sousa (2020 and 2022), the rule based on the symmetric around zero beta prior proposed by Sousa et al. (2020), the Large Posterior Mode (LPM) method by Cutillo et al. (2008), Amplitude-scale invariant Bayes Estimator (ABE) by Figueiredo and Nowak (2001) and Bayesian Adaptive Multiresolution Smoother (BAMS) by Vidakovic and Ruggeri (2001).

This paper is organized as follows: the statistical model for de aggregated curves and the estimation procedure of the component curves are defined in Section 2. The descriptions of the considered bayesian methods in the simulation studies are in Section 3. Section 4 is dedicated to the simulation studies results and analysis. We finish with further considerations and discussions in Section 5. 

\section{Statistical model and estimation procedure}
We consider a univariate function $A(t) \in \mathbb{L}_2(\mathbb{R})=\{f: \mathbb{R} \rightarrow \mathbb{R}|\int f^2 < \infty\}$ that can be written as
\begin{equation}\label{funmodel}
A(t) = \sum_{l=1}^{L}y_{l} \alpha_{l}(t) + e(t),
\end{equation}
where $\alpha_{l}(t) \in \mathbb{L}_2(\mathbb{R})$ are unknown component functions, $y_{l}$ are known real valued weights, $l=1, \cdots, L$, and $\{e(t),t \in \mathbb{R}\}$ is a zero mean gaussian process with unknown variance $\sigma^2$, $\sigma > 0$. The estimation of the functions $\alpha_l(t)$ is considered in this work. In fact, this estimation process is usually done by multivariate methods (Brown et al., 1998a,b) or functional data analysis (FDA) approach (Dias et al., 2009, 2013). In this last point of view, each function $\alpha_l$ of \eqref{funmodel} is represented in terms of some functional basis such as splines, B-splines or Fourier basis, for example. Here, we expand each component function by wavelet basis,
\begin{equation} \label{alphawav}
\alpha_l(t) = \sum_{j,k \in \mathbb{Z}} \gamma_{jk}^{(l)} \psi_{jk}(t), \hspace{1cm} l=1, \cdots, L,
\end{equation}
where $\{\psi_{jk}(x) = 2^{j/2} \psi(2^j x - k),j,k \in \mathbb{Z} \}$ is an orthonormal wavelet basis for $\mathbb{L}_2(\mathbb{R})$ constructed by dilations $j$ and translations $k$ of a function $\psi$ called wavelet or mother wavelet and $\gamma_{jk}^{(l)}$ 's are unknown wavelet coefficients of the expansion of the component function $\alpha_l$. Note that we consider the same wavelet family for all component functions expansion. Thus, the problem of estimating the functions $\alpha_l$ becomes a problem of estimating the finite number of wavelet coefficients $\gamma_{jk}^{(l)}$'s of the representation \eqref{alphawav}. Further, the magnitude of the wavelet coefficients allow to recover local features of the component functions, such as discontinuities points, spikes and oscillations, due the well localization in time and frequency domain of wavelets. This characteristic does not occur in spline-based and Fourier basis representations. 

In practice, one observes $I$ samples of the aggregated curve $A(t)$ at $M = 2^J$ equally spaced locations $t_1, \cdots, t_M$, i.e, our dataset is $\{(t_m, A_i(t_m))$, $m = 1, \cdots, M $ and $i = 1, \cdots, I\}$. Thus, the discrete version of \eqref{funmodel} is 
\begin{equation}\label{discmodel}
A_i(t_m) = \sum_{l=1}^{L}y_{il}\alpha_l(t_m) + e_i(t_m), \hspace{0.5cm} i=1,\cdots,I, \hspace{0.5cm} m=1,\cdots,M=2^J,
\end{equation}
where $e_i(t_m)$ are independent and identically normal distributed random noise with zero mean and variance $\sigma ^2$, $\forall i,m$. Further, the $I$ samples are obtained at the same locations $t_1, \cdots, t_M$ but the weights of the linear combinations are allowed to be different from one sample to another. We can rewrite \eqref{discmodel} in matrix notation as
\begin{equation}\label{vecmodel}
\boldsymbol{A} = \boldsymbol{\alpha}\boldsymbol{y} + \boldsymbol{e},
\end{equation}
where $\boldsymbol{A} = (A_{mi}=A_i(t_m))_{1 \leq m \leq M, 1 \leq i \leq I}$, $\boldsymbol{\alpha} = (\alpha_{ml}=\alpha_l(t_m))_{1 \leq m \leq M, 1 \leq l \leq L}$, $\boldsymbol{y} = (y_{li})_{1 \leq l \leq L, 1 \leq i \leq I}$ and $\boldsymbol{e} = (e_{mi}=e_i(t_m))_{1 \leq m \leq M, 1 \leq i \leq I}$.

The wavelet shrinkage procedure will be made in the wavelet domain to estimate the coefficients $\gamma$'s of \eqref{alphawav}. For this reason, we apply a discrete wavelet transform (DWT) on the original aggregated data, which can be represented by a $M \times M$ wavelet transformation matrix $\boldsymbol{W}$, and applied on both sides of \eqref{vecmodel}, i.e,
\begin{align}\label{matrixmodel}
\boldsymbol{W}\boldsymbol{A} &= \boldsymbol{W}(\boldsymbol{\alpha}\boldsymbol{y} + \boldsymbol{e}) \nonumber \\
\boldsymbol{W}\boldsymbol{A} &= \boldsymbol{W}\boldsymbol{\alpha}\boldsymbol{y} + \boldsymbol{W}\boldsymbol{e} \nonumber \\
\boldsymbol{D} &= \boldsymbol{\Gamma}\boldsymbol{y} + \boldsymbol{\varepsilon},
\end{align}
where $\boldsymbol{D} = \boldsymbol{W}\boldsymbol{A} = (d_{mi})_{1 \leq m \leq M, 1 \leq i \leq I}$ is the matrix with the empirical wavelet coefficients of the aggregated curves, $\boldsymbol{\Gamma} = \boldsymbol{W}\boldsymbol{\alpha} = (\gamma_{ml})_{1 \leq m \leq M, 1 \leq l \leq L}$ is the matrix with the unknown wavelet coefficients of the component curves and $\boldsymbol{\varepsilon} = \boldsymbol{W}\boldsymbol{e} =  (\varepsilon_{mi})_{1 \leq m \leq M, 1 \leq i \leq I}$ is the matrix with the random errors on the wavelet domain, which remain zero mean normal distributed with variance $\sigma^2$ due the orthogonality property of wavelet transforms. Thus, for a particular empirical wavelet coefficient $d_{mi}$ of $\boldsymbol{D}$, one has the additive model
\begin{equation} \label{coefmodel} 
d_{mi} = \sum_{l=1}^{L} y_{li}\gamma_{ml} + \varepsilon_{mi} = \theta_{mi} + \varepsilon_{mi},
\end{equation}
where $\theta_{mi} = \sum_{l=1}^{L} y_{li} \gamma_{ml}$ and $\varepsilon_{mi}$ is zero mean normal with variance $\sigma^2$, i.e, a single empirical wavelet coefficient of the aggregated curve is also a linear combination of the unknown wavelet coefficients of the component curves plus a random error. Moreover, the weights of this linear combination on wavelet domain remain the same that the original combination of the curves at time domain. 

The estimation of the wavelet coefficients matrix $\boldsymbol{\Gamma}$  in \eqref{matrixmodel} is done by applying a wavelet shrinkage rule $\delta$ on each single empirical wavelet coefficient $d$, obtaining the matrix $\boldsymbol{\delta(\boldsymbol{D})}$ such that
\begin{equation}
\boldsymbol{\delta(\boldsymbol{D})} = (\delta(d_{mi}))_{1 \leq m \leq M, 1 \leq i \leq I}.
\end{equation}
We can see the matrix $\boldsymbol{\delta(\boldsymbol{D})}$ as a denoising version of $\boldsymbol{D}$, i.e, the shrinkage rule $\delta(d)$ acts by denoising the empirical coefficient $d$ in order to estimate $\theta$ in \eqref{coefmodel}, $\delta(d) = \hat{\theta}$. Thus, the estimation $\boldsymbol{\hat{\Gamma}}$ of the wavelet coefficients matrix $\boldsymbol{\Gamma}$ is given by least squares method,
\begin{equation}
\boldsymbol{\hat{\Gamma}} = \boldsymbol{\delta(\boldsymbol{D})}\boldsymbol{y^{t}}\boldsymbol{(yy^{t})^{-1}},
\end{equation}
and finally $\boldsymbol{\alpha}$ can be estimated at locations $t_1, \cdots, t_M$ by the inverse discrete wavelet transformation (IDWT),
\begin{equation}
\boldsymbol{\hat{\alpha}} = \boldsymbol{W^t}\boldsymbol{\hat{\Gamma}}.
\end{equation}

For more details about wavelet shrinkage in aggregated curves, see Sousa (2022).  

\section{Bayesian methods}
There are several available wavelet shrinkage methods in the literature. Most of them are thresholding, i.e, the rule shrinks sufficient small empirical wavelet coefficients as exactly zero. Two extremely applied thresholding rules are the so called hard and soft rules proposed by Donoho and Johnstone (1994b). In this work will be considered bayesian wavelet shrinkage methods that have been proposed in recent years. In general, these methods propose a prior distribution to the wavelet coefficients and estimate them according to a loss function, as quadratic loss function for example, by the Bayes rule. The main advantage of bayesian methods is the ability to incorporate prior information about the wavelet coefficients, as sparsity, boundedness, self similarity and others by convenient choices of prior distributions and their hyperparameters values. 
       
In the next subsections, we briefly describe the considered bayesian shrinkage rules in this work.
\subsection{Shrinkage rule under logistic prior}
The shrinkage rule under logistic prior was proposed by Sousa (2020). It assumes a mixture of a point mass function at zero and a symmetric logistic distribution as prior to a single linear combination of wavelet coefficients $\theta$,
\begin{equation}\label{prior}
\pi(\theta;p,\tau) = p \delta_{0}(\theta) + (1-p)g(\theta;\tau),
\end{equation}
where $p \in (0,1)$, $\delta_{0}(\theta)$ is the point mass function at zero and $g(\theta;\tau)$ is the logistic density function symmetric around zero, for $\tau > 0$, 
\begin{equation} \label{log}
g(\theta;\tau) = \frac{\exp\{-\frac{\theta}{\tau}\}}{\tau(1+\exp\{-\frac{\theta}{\tau}\})^2}\mathbb{I}_{\mathbb{R}}(\theta).
\end{equation} 
Under squared loss function, the associated bayesian shrinkage rule is the posterior expected value of $\theta$, $\mathbb{E}_{\pi}(\theta|d)$, that under the model prior \eqref{prior}, is given by (Sousa, 2020),
\begin{equation}\label{rule}
\delta(d) = \mathbb{E}_{\pi}(\theta|d) = \frac{(1-p)\int_\mathbb{R}(\sigma u + d)g(\sigma u +d ; \tau)\phi(u)du}{\frac{p}{\sigma}\phi(\frac{d}{\sigma})+(1-p)\int_\mathbb{R}g(\sigma u +d ; \tau)\phi(u)du},
\end{equation}
where $\phi(\cdot)$ is the standard normal density function. The shrinkage rule \eqref{rule} under the model \eqref{log} is called logistic shrinkage rule and has interesting features under estimation point of view. First, its hyperparameters $p$ and $\tau$ control the degree of shrinkage of the rule. Higher values of $\tau$ or $p$ imply higuer shrinkage level, i.e, the rule will reduce severely the magnitudes of the empirical coefficients. Further, as described in Sousa (2020), the logistic shrinkage rule had good performances in terms of averaged mean squared error in simulation studies against standard shrinkage or thresholding procedures.

\subsection{Shrinkage rule under beta prior}
Sousa et al. (2020) proposed the use of a mixture of a point mass function at zero and the beta distribution with symmetric support around zero as a prior distribution to the wavelet coefficients,
\begin{equation}\label{priorbeta}
\pi(\theta;p,a,m) = p \delta_{0}(\theta) + (1-p)g(\theta;a,m),
\end{equation}
and
\begin{equation}\label{eq:beta}
g(\theta;a,m) = \frac{(m^2 - \theta^2)^{(a-1)}}{(2m)^{(2a-1)}B(a,a)}\mathbb{I}_{[-m,m]}(\theta),
\end{equation}
\noindent where $B(\cdot , \cdot)$ is the standard beta function, $a>0$ and $m>0$ are the parameters of the distribution, and $\mathbb{I}_{[-m,m]}(\cdot)$ is an indicator function equal to 1 for its argument in the interval $[-m,m]$ and 0 else.

 For $a>1$, the density function (\ref{eq:beta}) is unimodal around zero and as $a$ increases, the density becomes more concentrated around zero. This is an important feature for wavelet shrinkage methods, since high values of $a$ imply higher levels of shrinkage in the empirical coefficients, which results in sparse estimated coefficients. Density \eqref{eq:beta} becomes uniform for $a=1$, which was already considered by Angelini and Vidakovic (2004) as prior to the wavelet coefficients.
 
 Based on the fact that much of the noise information present in the data can be obtained on the finer resolution scale, for the robust $\sigma$ estimation, Donoho and Johnstone (1994a) suggest

\begin{equation}\label{eq:sigma}
\hat{\sigma} = \frac{\mbox{median}\{|d_{J-1,k}|:k=0,...,2^{J-1}\}}{0.6745}.
\end{equation}
Angelini and Vidakovic (2004) suggest the hyperparameters $p$ and $m$ be dependent on the level of resolution $j$ according to the expressions
\begin{equation}\label{eq:alpha}
p = p(j) = 1 - \frac{1}{(j-J_{0}+1)^\gamma}
\end{equation}
and
\begin{equation}\label{eq:m}
m = m(j) = \max_{k}\{|d_{jk}|\},
\end{equation}
where $J_ 0 \leq j \leq J-1$, $J_0$ is the primary resolution level and $\gamma > 0$. They also suggest that in the absence of additional information, $\gamma = 2$ can be adopted.

\subsection{Large Posterior Mode (LPM)}
Cutillo et al. (2008) proposed a bayesian thresholding rule that is based on the Maximum a Posteriori (MAP) principle. Under the model \eqref{coefmodel} with gaussian noise, it is assumed a normal prior for $\theta$, i.e, $\theta|t^2 \sim \mathrm{N}(0,t^2)$, where $t^2 \sim (t^2)^{-k}$, $k > 1/2$. The LPM thresholding rule picks the larger mode in absolute value of the posterior distribution of $\theta|d$. Further, there is an interesting feature of the posterior that allows the Bayes rule to be thresholding. The posterior can be unimodal at zero or bimodal trivially at zero and at another local mode. For the unimodal case, the empirical coefficient is then shrunk to zero. For the second one, it is shrunk slightly to the local mode. The closed form of the rule is
\begin{equation}
\delta_{\mathrm{LMP}}(d) = \frac{d + sgn(d)\sqrt{d^2 - 4\sigma^2 (2k-1)}}{2} \mathbb{I}_{[\lambda_{\mathrm{LPM}},+\infty)}(|d|),
\end{equation}
where $\lambda_{\mathrm{LPM}} = 2\sigma \sqrt{2k-1}$.

\subsection{Amplitude-Scale Invariant Bayes Estimator (ABE)}
Figueiredo and Nowak (2001) proposed a bayesian thresholding rule that does not depend on prior hiperparameters, which must be elicited for other bayesian shrinkage/thresholding methods. It assumes an amplitude-scale invariant (noninformative) prior $\pi(\theta) \propto |\theta|^{-1}$ for $\theta$ from the  model \eqref{coefmodel}. The Bayes rule is thresholding, given by
\begin{equation}
\delta_{\mathrm{ABE}}(d) = \frac{(d^2 - 3\sigma^2)_{+}}{d},
\end{equation} 
where $(x)_{+} = \max\{0,x\}$. Note that the rule depends only on the noise variance parameter $\sigma^2$.

\subsection{Bayesian Adaptive Multiresolution Smoother (BAMS)}
BAMS method was proposed by Vidakovic and Ruggeri (2001) and is obtained by considering $\sigma^2 \sim \mathrm{Exp}(1/\mu)$, $\mu >0$, and the following prior distribution $\pi$ to a single wavelet coefficient,
\begin{equation}
\pi(\theta) = \alpha \delta_0(\theta) + (1-\alpha)\mathrm{DE}(0,\tau),
\end{equation}
where $\alpha \in (0,1)$, $\delta_0(\cdot)$ is a point mass function at zero and $\mathrm{DE}(0,\tau)$ is the double exponential density with location parameter equals to zero and scale parameter $\tau$, $\tau >0$, given by $g(\theta;0,\tau) = \frac{1}{2\tau}\exp\Big\{-\frac{|\theta|}{\tau}\Big\}$, $\theta \in \mathbb{R}$. The associated shrinkage rule under squared error loss, $\delta_{\mathrm{BAMS}}(d) = \mathbb{E}(\theta|d)$ is  
\begin{equation}
\delta_{\mathrm{BAMS}}(d) = \frac{(1-\alpha) m(d) \delta(d)}{(1-\alpha) m(d) + \alpha \mathrm{DE}(0,\frac{1}{\sqrt{2\mu}})},
\end{equation}
where $m(\cdot)$ and $\delta(\cdot)$ are the predictive distribution of $d$ and the shrinkage rule respectively under assumption that $\theta \sim \mathrm{DE}(0,\tau)$ and given by  
\begin{equation}
m(d) = \frac{\tau \exp \Big\{ -\frac{|d|}{\tau} \Big \} - \frac{1}{\sqrt{2\mu}}\exp\{-\sqrt{2\mu}|d|\}}{2\tau^2 - \frac{1}{\mu}} \nonumber
\end{equation}
and
\begin{equation}
\delta(d) = \frac{\tau \left(\tau^2 - \frac{1}{2\mu}\right)d\exp\{-|d|/\tau\} + \frac{\tau^2}{\mu}\left( \exp\{-|d|/\sqrt{2\mu}\} - \exp\{-|d|/\tau\} \right)}{\left(\tau^2 - \frac{1}{2\mu} \right)\left( \tau \exp\{ -|d|/\tau \} - \frac{1}{\sqrt{2\mu}} \exp\{-|d|/\sqrt{2\mu} \}\right)}. \nonumber
\end{equation}

\begin{figure}[H]
\centering
\includegraphics[scale=1.0]{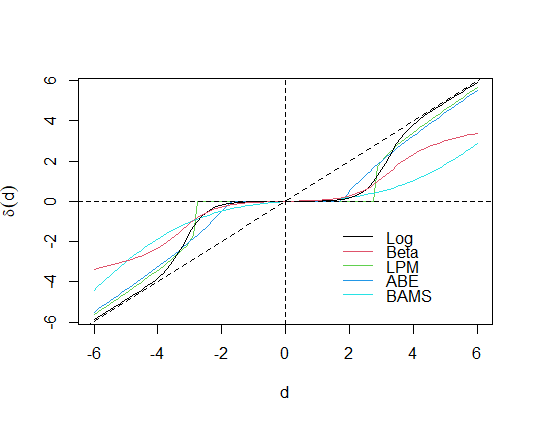}
\caption{Logistic, Beta, LPM, ABE and BAMS bayesian rules. The first one is a shrinkage rule while LPM and ABE are thresholding rules.}\label{fig:rules}
\end{figure}

Figure \ref{fig:rules} shows BAMS, LPM and ABE rules. It is clear the shrinkage behaviour of BAMS, i.e, although it shrinks the empirical coefficient magnitude, the shrunk coefficient is not necessarily zero. On the other hand, LPM and ABE rules are thresholding, once they shrink an sufficiently small coefficient to zero. 

\section{Simulation studies}

\begin{figure}[H]
\centering
\subfigure[Bumps.]{
\includegraphics[scale=0.35]{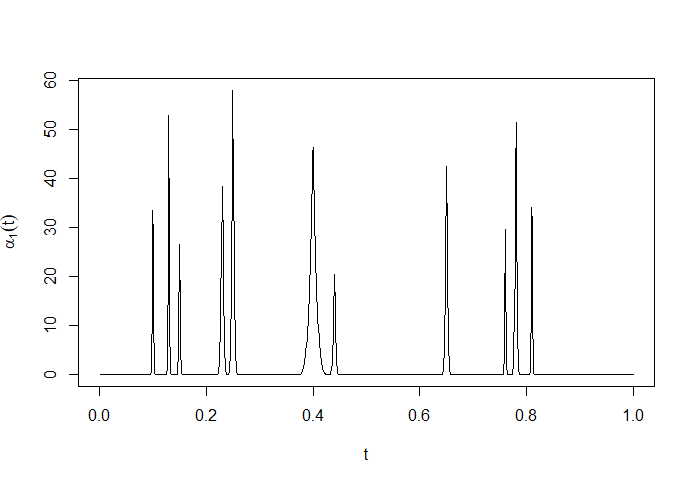}}
\subfigure[Blocks.]{
\includegraphics[scale=0.35]{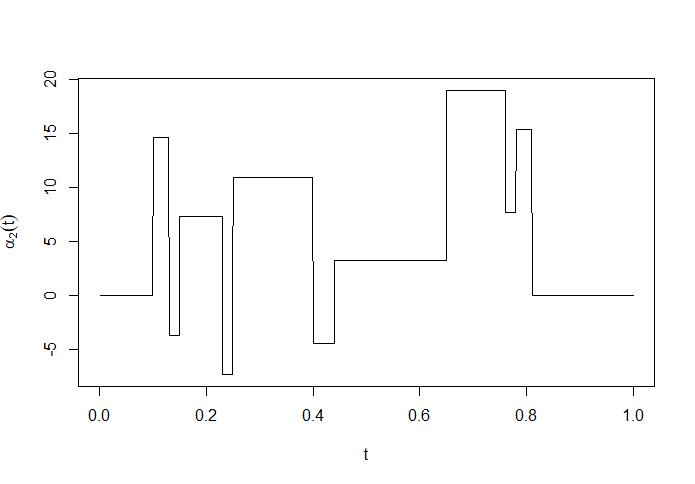}}
\subfigure[Doppler.]{
\includegraphics[scale=0.35]{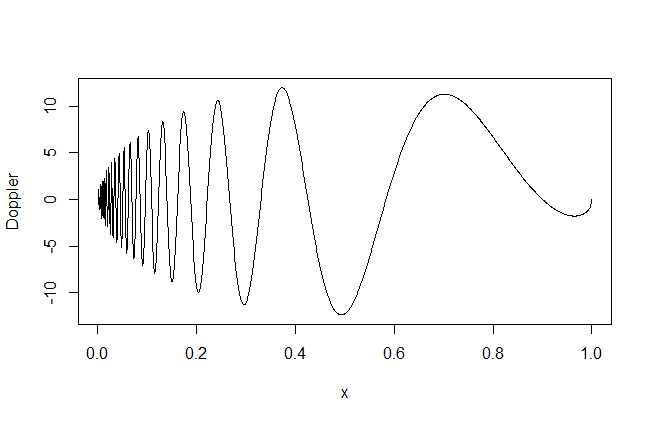}}
\subfigure[Heavisine.]{
\includegraphics[scale=0.35]{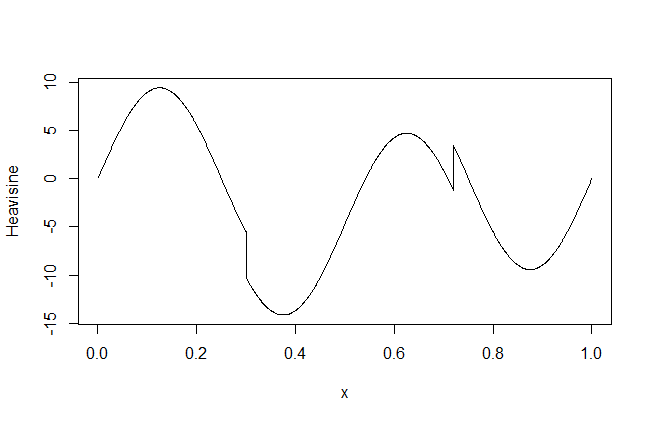}}
\subfigure[Logit.]{
\includegraphics[scale=0.35]{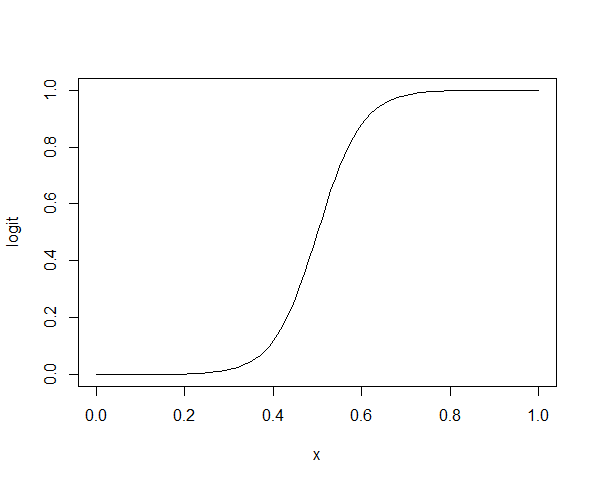}}
\subfigure[SpaHet.]{
\includegraphics[scale=0.35]{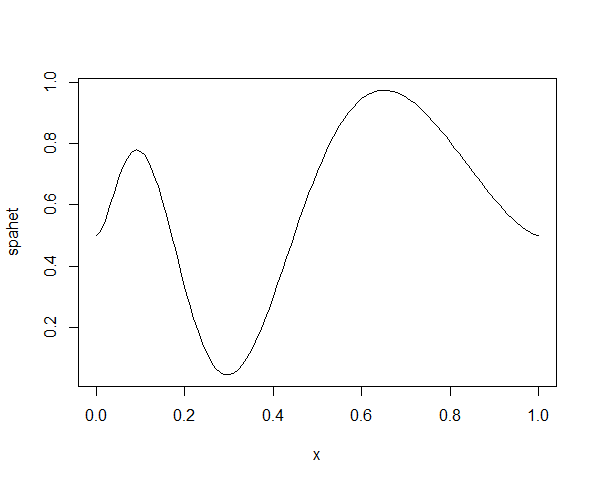}}
\caption{Functions used as component functions in the simulation studies.} \label{fig:functions}
\end{figure}

Simulation studies were conducted to evaluate the performances of the considered bayesian rules in estimating component curves. Six functions on $[0,1]$ were selected as component curves: the so called Donoho-Johnstone test functions Bumps, Blocks, Doppler and Heavisine and the functions Logit and SpaHet, that were used in simulation studies of Wand (2000), Ruppert (2003) and Goepp et al. (2018). The mathematical expressions of the functions are available in Table \ref{tab:compfun} and their plots are in Figure \ref{fig:functions}. The four Donoho-Johnstone test functions have interesting local features to be recovered, such as peaks (Bumps), oscillations (Doppler) and discontinuities (Blocks and Heavisine). On the other hand, Logit and SpaHet are smooth functions in the whole domain $[0,1]$.  

Three simulation studies were considered according to different number of component functions. Simulation 1 was conducted by considering two component functions ($L=2$), Bumps and Blocks. The Simulation 2 considered four component functions ($L=4$), Bumps, Blocks, Doppler and Logit. Finally, Simulation 3 considered the six component functions ($L=6$). For each simulation study, $I = 50$ aggregated curves were generated from model \eqref{discmodel} for $M=512=2^9$ and $1024=2^{10}$ equally spaced points on $[0,1]$. The zero mean gaussian noises were generated with variances according to signal to noise ratio (SNR) values $\mathrm{SNR}=  3$ and $9$. Smaller SNR values implies higuer amount of noise in the data. For each scenario of $M$ and SNR values, $N = 100$ replicates were done and the component functions points were estimated for each considered bayesian rule. The mean squared error (MSE) of the j-the replication of a component function $\alpha_l(t)$, 
\begin{equation}
\mathrm{MSE_l^{(j)}} = \frac{1}{M} \sum_{i=1}^{M}[{\hat \alpha_l^{(j)}(t_i)} - \alpha_l(t_i)]^2, \nonumber
\end{equation}
was calculated for each replicate and the averaged mean squared error (AMSE), 
\begin{equation}
\mathrm{AMSE_l} = \frac{1}{N} \sum_{j=1}^{N}\mathrm{MSE}_l^{(j)}, \nonumber
\end{equation}
was considered as performance measure, for each component function $\alpha_l(t)$, $l = 1,\cdots,L$. In all generated dataset, the DWT under Daubechies basis with ten null moments was applied to obtain the empirical coefficients vector.

\begin{table}[H]
\scalefont{0.80}
\centering
\label{my-label}
\begin{tabular}{|c|}
\hline
\textbf{BUMPS} \\ 
$ f(x) = \sum_{l=1}^{11} h_l K\left(\frac{x - x_l}{w_l} \right)$ \\
where \\

$K(x) = (1 + |x|)^{-4}$ \\

$(x_l)_{l=1}^{11} = (0.1, 0.13, 0.15, 0.23, 0.25, 0.40, 0.44, 0.65, 0.76, 0.78, 0.81)$\\

$(h_l)_{l=1}^{11} = (4, 5, 3, 4, 5, 4.2, 2.1, 4.3, 3.1, 5.1, 4.2)$  \\

$(w_l)_{l=1}^{11} = (0.005, 0.005, 0.006, 0.01, 0.01, 0.03, 0.01, 0.01, 0.005, 0.008, 0.005)$\\ \hline 

\textbf{BLOCKS} \\
$ f(x) = \sum_{l=1}^{11} h_l K(x - x_l)$ \\
where \\

$K(x) = (1 + \mathrm{sgn}(x))/2$ \\

$(x_l)_{l=1}^{11} = (0.1, 0.13, 0.15, 0.23, 0.25, 0.40, 0.44, 0.65, 0.76, 0.78, 0.81)$  \\

$(h_l)_{l=1}^{11} = (4, -5, 3, -4, 5, -4.2, 2.1, 4.3, -3.1, 2.1, -4.2)$ \\ \hline 

\textbf{DOPPLER} \\
$ f(x) = \sqrt{x(1-x)}\sin\left(\frac{2.1 \pi}{x + 0.05} \right)$ \\ \hline 

\textbf{HEAVISINE} \\
$ f(x) = 4\sin(4 \pi x) - \mathrm{sgn}(x - 0.3) - \mathrm{sgn}(0.72 - x)$ \\ \hline

\textbf{LOGIT} \\ 
$f(x) = \frac{1}{1+\exp\{-20(x-0.5)\}} $ \\ \hline

\textbf{SPAHET}  \\ 
$f(x) = \sqrt{x(1-x)}\sin\left(\frac{2\pi(1+2^{-0.6})}{x+2^{-0.6}}\right)$ \\ \hline

\end{tabular}
\caption{Test functions definitions for $x \in [0,1]$ used as component functions in the simulation studies.}\label{tab:compfun}
\end{table} 

\subsection{Simulation study 1}

\begin{figure}[H]
\centering
\subfigure[Ten generated samples.]{
\includegraphics[scale=0.55]{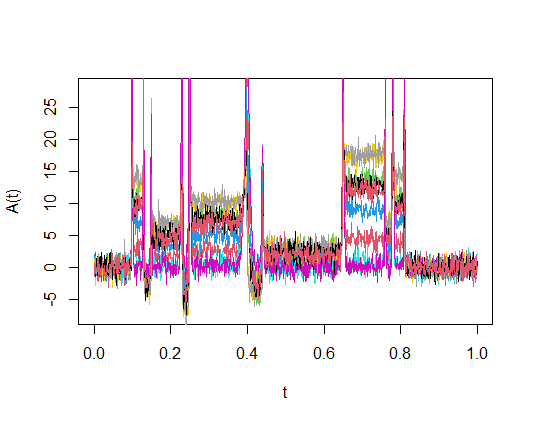}}
\subfigure[Empirical wavelet coefficients.]{
\includegraphics[scale=0.55]{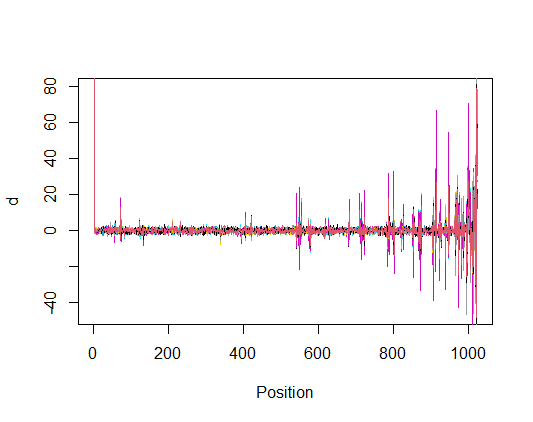}}
\subfigure[Boxplot - bumps (SNR = 3).]{
\includegraphics[scale=0.55]{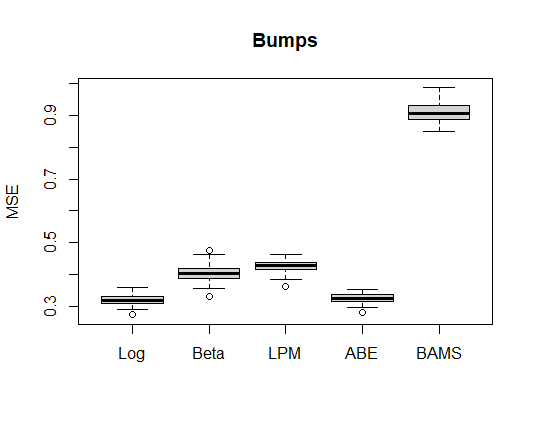}}
\subfigure[Boxplot - blocks (SNR = 3).]{
\includegraphics[scale=0.55]{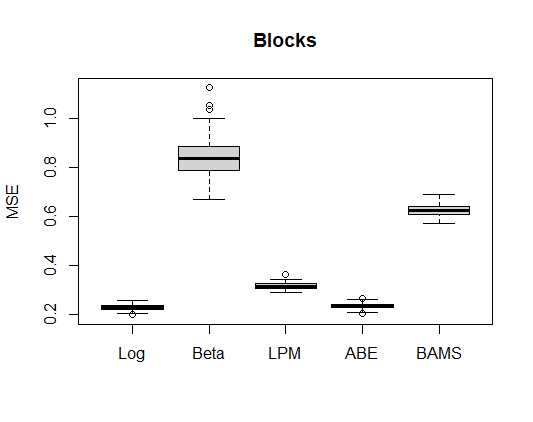}}
\subfigure[Boxplot - bumps (SNR = 9).]{
\includegraphics[scale=0.55]{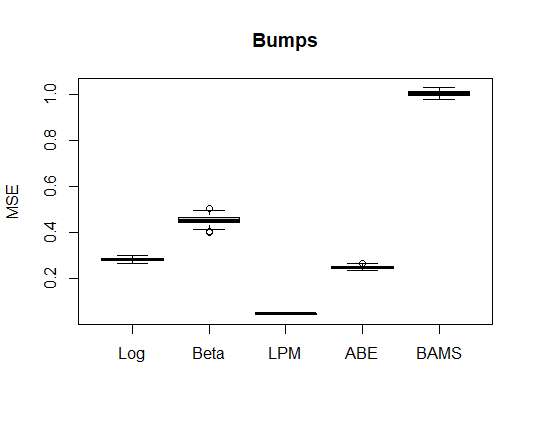}}
\subfigure[Boxplot - blocks (SNR = 9).]{
\includegraphics[scale=0.55]{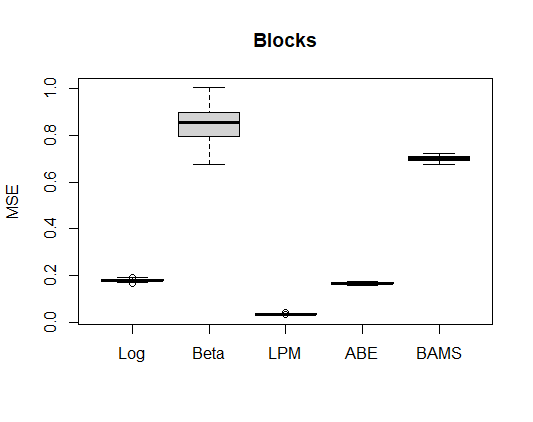}}
\caption{Ten samples of Simulation 1 (a) and their empirical wavelet coefficients (b). Boxplots of rules MSE for bumps (c) and blocks (d) component functions for $M=1024$ and $\mathrm{SNR}=3$ and bumps (e) and blocks (f) for $\mathrm{SNR}=9$.} \label{fig:sim1}
\end{figure}

In this simulation study, the $50$ samples were generated by aggregation of Bumps and Blocks functions with different weights for each function in the linear combination of model \eqref{discmodel}. Figures \ref{fig:sim1} (a) and (b) show ten generated samples and their empirical wavelet coefficients respectively for $M = 1024$ and $\mathrm{SNR} = 3$. It is possible to observe local features of both component functions in the samples, i.e., the spikes of Bumps, discontinuities and piecewise constant parts of Blocks. 

Table \ref{tab:amse1} show the obtained AMSEs (with standard deviations of MSE) of the bayesian rules for each scenario of $M$ and SNR. The lowest AMSEs are in bold. In fact, the shrinkage rule under logistic prior was the best for all the scenarios with $\mathrm{SNR} = 3$ for both component functions, while LPM rule was the best one for the scenarios with $\mathrm{SNR} = 9$. The number of sample points $M$ did not have meaningful impact in the estimation process but the rules performed better for $\mathrm{SNR} = 9$ as expected. Figures \ref{fig:sim1} (c) and (d) present the boxplots of MSEs for Bumps and Blocks component functions respectively for $M = 1024$ and $\mathrm{SNR} = 3$ and (e) and (f) for $\mathrm{SNR}=9$ scenarios. Note that BAMS and shrinkage rule under beta prior did not have good performances in comparison with the other rules while ABE rule had a good performance, very close to shrinkage rule under logistic prior for both component functions. Similar results were obtained for $M = 512$. 

\begin{table}[H]
\scalefont{0.7}
\centering
\label{my-label}
\begin{tabular}{|c|c|c|c||c|c|c|c|}
\hline
 \multicolumn{6}{|c|}{\textbf{Simulation study 1} - $\boldsymbol{L=2}$} \\ \hline
  \multicolumn{2}{|c|}{} & \multicolumn{2}{|c||}{$\boldsymbol{M=512}$} & \multicolumn{2}{|c|}{$\boldsymbol{M=1024}$}\\ \hline
\textbf{Function}  & \textbf{Method} & \textbf{SNR = 3} & \textbf{SNR = 9} & \textbf{SNR = 3} & \textbf{SNR = 9} \\ \hline \hline

Bumps&     LOG     &       \textbf{0.2932  (0.0187)}       &       0.3310  (0.0103) 								    &       \textbf{0.3214  (0.0158)}       &       0.2827  (0.0077) \\
       &BETA   &       0.3761  (0.0271)        &       0.4671  (0.0186) & 								         0.4044  (0.0259)        &       0.4556  (0.0211) \\
   &LPM    &       0.3083  (0.0170)        &       \textbf{0.0342  (0.0019)} 								&      0.4266  (0.0172)        &       \textbf{0.0471  (0.0020)} \\
    &ABE    &       0.3091  (0.0195)        &       0.3016  (0.0088) 						 &       0.3267  (0.0152)        &       0.2493  (0.0059) \\
      &BAMS   &       11.2350 (0.0411)        &       12.7690 (0.0160) &      0.9092  (0.0300)        &       10.061  (0.0112) \\ \hline

Blocks   &     LOG     &       \textbf{0.2711  (0.0169)}       &       0.2643  (0.0087) 						   &       \textbf{0.2275  (0.0114)}       &       0.1796  (0.0046) \\
       &BETA   &       0.8775  (0.0699)        &       0.9446  (0.0664)  &       0.8391  (0.0797)        &       0.8505  (0.0712) \\
       &LPM    &       0.3045  (0.0188)        &       \textbf{0.0340  (0.0022)}  &       0.3151  (0.0130 )&      \textbf{0.0349  (0.0016)} \\
       &ABE    &       0.2853  (0.0175)        &       0.2525  (0.0083)  &       0.2331  (0.0111)        &       0.1669  (0.0039) \\
       &BAMS   &       0.9784  (0.0362)        &       11.2562 (0.0155) &       0.6231  (0.0252)        &       0.7002  (0.0095) \\ \hline

\end{tabular}
\caption{AMSE (standard deviation) of simulation study 1 for aggregated data generated with the component functions Bumps and Blocks.}\label{tab:amse1}
\end{table} 

\subsection{Simulation study 2}

\begin{figure}[H]
\centering
\subfigure[Ten generated samples.]{
\includegraphics[scale=0.55]{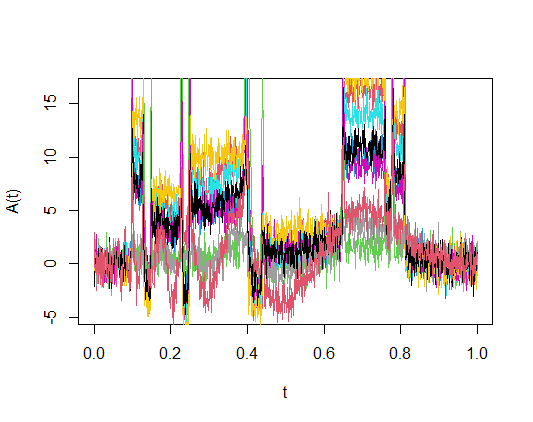}}
\subfigure[Empirical wavelet coefficients.]{
\includegraphics[scale=0.55]{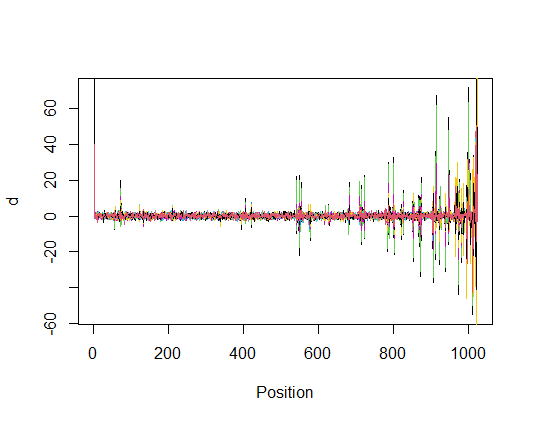}}
\subfigure[Boxplot - bumps.]{
\includegraphics[scale=0.55]{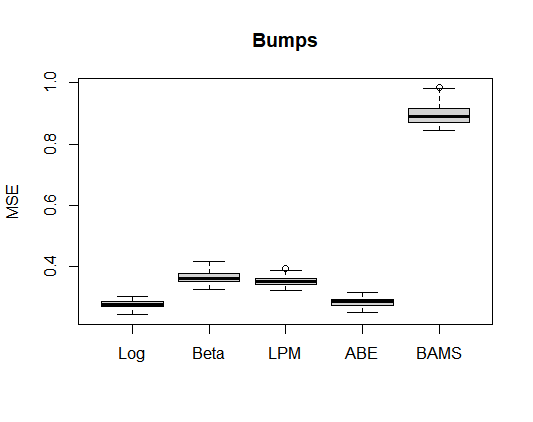}}
\subfigure[Boxplot - blocks.]{
\includegraphics[scale=0.55]{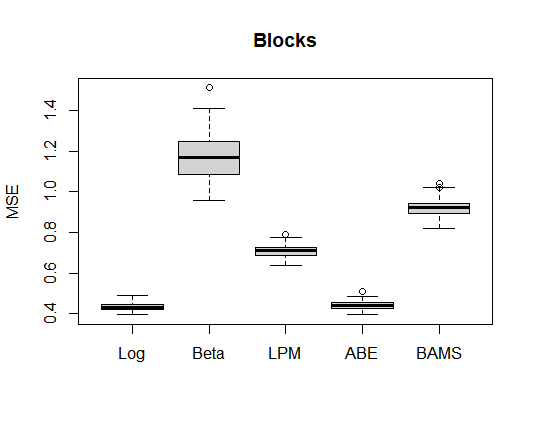}}
\subfigure[Boxplot - doppler.]{
\includegraphics[scale=0.55]{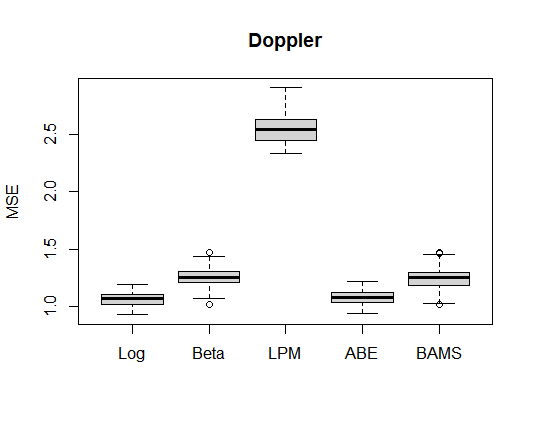}}
\subfigure[Boxplot - logit.]{
\includegraphics[scale=0.55]{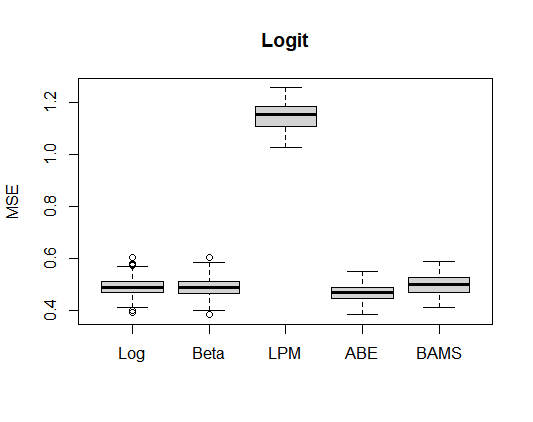}}
\caption{Ten samples of Simulation 2 (a) and their empirical wavelet coefficients (b). Boxplots of rules MSE for bumps (c), blocks (d), doppler (e) and logit (f) component functions for $M=1024$ and $\mathrm{SNR}=3$.} \label{fig:sim2}
\end{figure}

In this study, 50 samples were generated according to model \eqref{discmodel} by considering Bumps, Blocks, Doppler and Logit as underlying component functions ($L=4$). Figures \ref{fig:sim2} (a) and (b) show ten generated samples and their respective empirical wavelet coefficients. Note that it is more difficult to visualize local features of the component functions in the aggregated data, but a good estimation process should recover the spikes of Bumps, the discontinuities and piecewise constant parts of Blocks, the oscillations of Doppler and the smoothness of Logit. 

Table \ref{tab:amse2} presents the AMSEs and their standard deviations of Simulation 2. For the scenarios under SNR = 9, LPM rule had the best performance for the four component functions and both $M$ values, as occurred in Simulation study 1. For SNR = 3, the shrinkage rule under logistic prior and ABE had the best performances, the first one was the best for Bumps and Blocks functions and the second one the best for Doppler and Logit functions, but their overall works were very close with each other, which is important, since one rule is applied for estimating the all the component functions in practice.    

Figures \ref{fig:sim1} (c)-(f) show the boxplots of MSE for Bumps, Blocks, Doppler and Logit functions respectively for $M = 1024$ and SNR = 3. Note again that BAMS and the shrinkage rule under beta prior did not have good performance for the four underlying functions.

\begin{table}[H]
\scalefont{0.7}
\centering
\label{my-label}
\begin{tabular}{|c|c|c|c||c|c|c|c|}
\hline
 \multicolumn{6}{|c|}{\textbf{Simulation study 2} - $\boldsymbol{L=4}$} \\ \hline
  \multicolumn{2}{|c|}{} & \multicolumn{2}{|c||}{$\boldsymbol{n=512}$} & \multicolumn{2}{|c|}{$\boldsymbol{n=1024}$}\\ \hline
\textbf{Function}  & \textbf{Method} & \textbf{SNR = 3} & \textbf{SNR = 9} & \textbf{SNR = 3} & \textbf{SNR = 9} \\ \hline \hline

Bumps&LOG	&	\textbf{0.3653	(0.0232)}	&	0.4129	(0.0134)	&	\textbf{0.2776}	(0.0133)	&	0.2736	(0.0070)\\
&BETA	&	0.4686	(0.0351)	&	0.5709	(0.0199)	&	0.3648	(0.0200)	&	0.4265	(0.0163)\\
&LPM	&	0.4021	(0.0275)	&	\textbf{0.0449	(0.0028)}	&	0.3536	(0.0144)	&	\textbf{0.0392	(0.0017)}\\
&ABE	&	0.3785	(0.0234)	&	0.3643	(0.0115)	&	0.2848	(0.0138)	&	0.2430	(0.0056)\\
&BAMS	&	13.432	(0.0557)	&	14.910	(0.0206)	&	0.8957	(0.0279)	&	0.9937	(0.0112)\\ \hline

Blocks & LOG	&	\textbf{0.4702	(0.0370)}	&	0.3657	(0.0141)	&	\textbf{0.4336	0.0199}	&	0.2702	(0.0092) \\
&BETA	&	11.348	(0.1024)	&	11.184	(0.0917)	&	11.769	(0.1066)	&	10.870	(0.1029)\\
&LPM	&	0.5930	(0.0377)	&	\textbf{0.0660	(0.0038)}	&	0.7112	(0.0309)	&	\textbf{0.0786	(0.0031)}\\
&ABE	&	0.4821	(0.0377)	&	0.3474	(0.0123)	&	0.4411	(0.0200)	&	0.2562	(0.0080)\\
&BAMS	&	13.636	(0.0813)	&	14.666	(0.0253)	&	0.9203	(0.0429)	&	0.9474	(0.0158)\\ \hline

Doppler&	LOG	&	13.489	(0.1123)	&	0.7043	(0.0411)	&	\textbf{10.592	(0.0623)}	&	0.4292	(0.0218)\\
&	BETA	&	19.660	(0.1966)	&	14.773	(0.1358)	&	12.548	(0.0873)	&	0.7968	(0.0609)\\
&	LPM	&	26.068	(0.1635)	&	\textbf{0.2925	(0.0168)}	&	25.441	(0.1104)	&	\textbf{0.2829	(0.0127)}\\
&	ABE	&	\textbf{13.015	(0.0995)}	&	0.5585	(0.0319)	&	10.732	(0.0616)	&	0.3584	(0.0173)\\
&	BAMS	&	15.768	(0.1279)	&	13.576	(0.0378)	&	12.383	(0.0923)	&	10.250	(0.0288)\\ \hline

Logit&	LOG	&	0.9529	(0.0905)	&	0.4387	(0.0309)	&	0.4866	(0.0399)	&	0.1840	(0.0110)\\
&	BETA	&	0.9929	(0.1061)	&	0.5040	(0.0408)	&	0.4858	(0.0427)	&	0.2250	(0.0172)\\
&	LPM	&	16.164	(0.1065)	&	\textbf{0.1785	(0.0104)}	&	11.475	(0.0495)	&	\textbf{0.1269	(0.0063)}\\
&	ABE	&	\textbf{0.8913	(0.0816)}	&	0.3554	(0.0217)	&	\textbf{0.4660	(0.0335)}	&	0.1453	(0.0081)\\
&	BAMS	&	15.619	(0.1488)	&	13.097	(0.0463)	&	0.4981	(0.0415)	&	0.3333	(0.0149)\\ \hline

\end{tabular}
\caption{AMSE (standard deviation) of simulation study 2 for aggregated data generated with the component functions Bumps, Blocks, Doppler and Logit.}\label{tab:amse2}
\end{table} 

\subsection{Simulation study 3}

\begin{figure}[H]
\centering
\subfigure[Ten generated samples.]{
\includegraphics[scale=0.4]{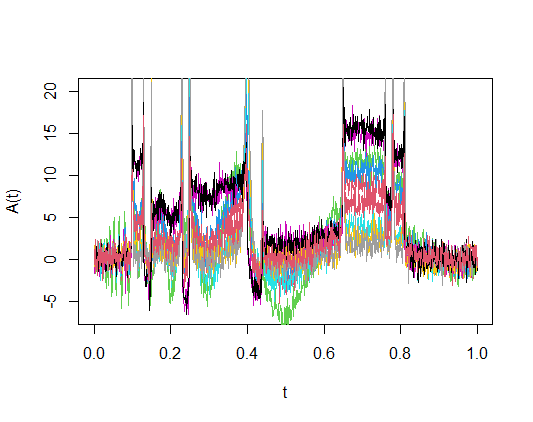}}
\subfigure[Empirical wavelet coefficients.]{
\includegraphics[scale=0.4]{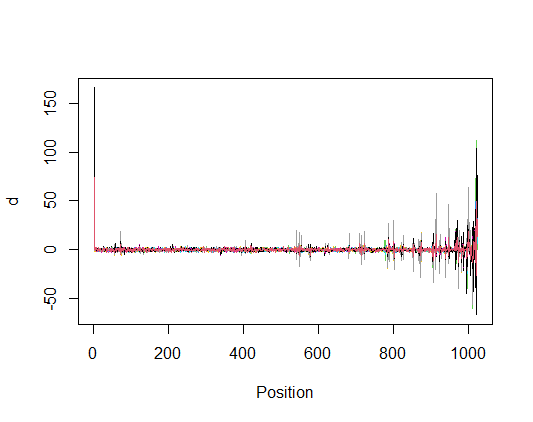}}
\subfigure[Boxplot - bumps.]{
\includegraphics[scale=0.4]{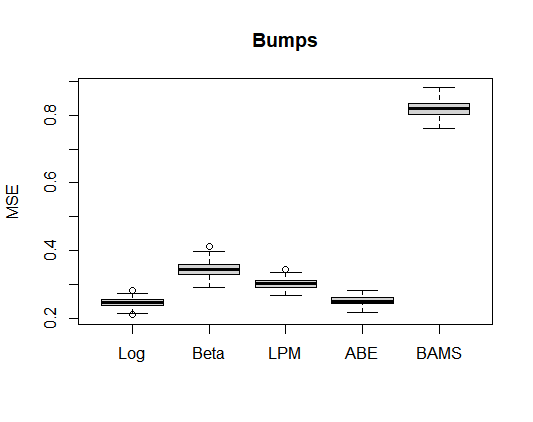}}
\subfigure[Boxplot - blocks.]{
\includegraphics[scale=0.4]{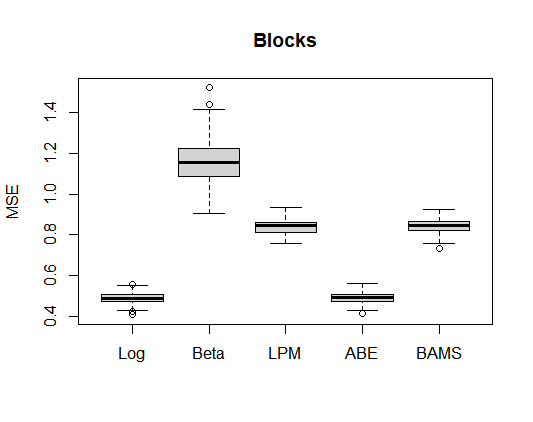}}
\subfigure[Boxplot - doppler.]{
\includegraphics[scale=0.4]{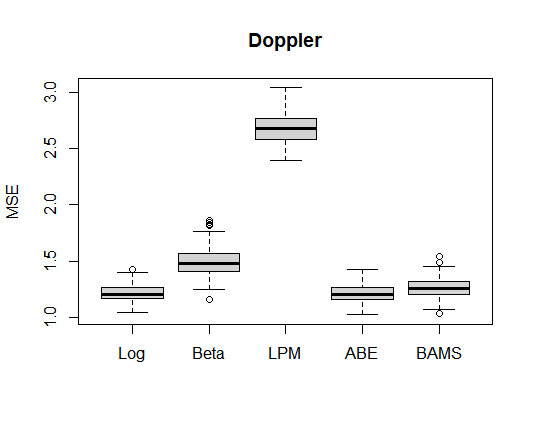}}
\subfigure[Boxplot - heavisine.]{
\includegraphics[scale=0.4]{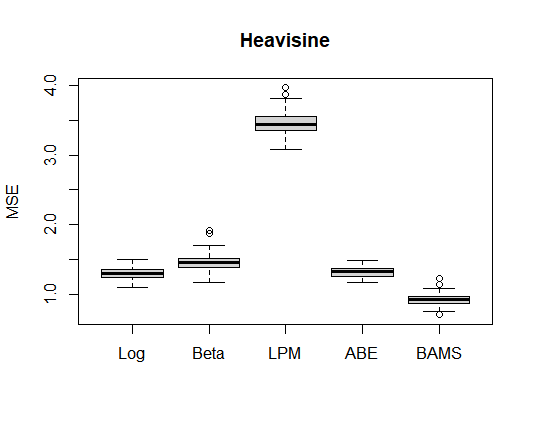}}
\subfigure[Boxplot - logit.]{
\includegraphics[scale=0.4]{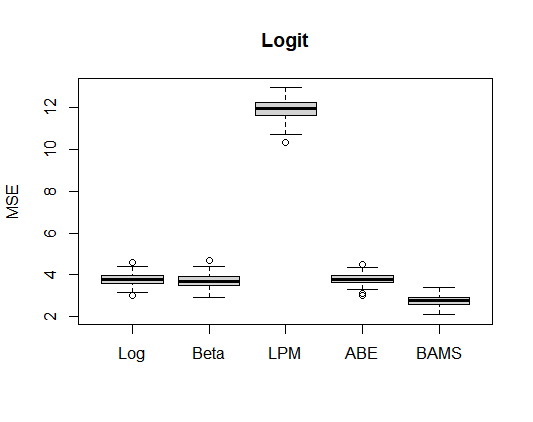}}
\subfigure[Boxplot - spahet.]{
\includegraphics[scale=0.4]{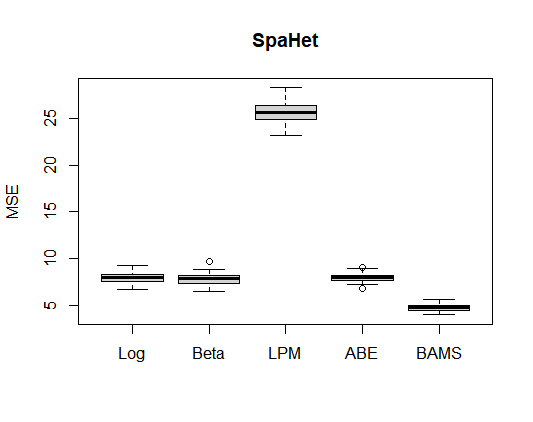}}
\caption{Ten samples of Simulation 3 (a) and their empirical wavelet coefficients (b). Boxplots of rules MSE for bumps (c), blocks (d), doppler (e), heavisine (f), logit (g) and spahet (h) component functions for $M=1024$ and $\mathrm{SNR}=3$.} \label{fig:sim3}
\end{figure}

Finally, Simulation study 3 considered all the six functions of Table \ref{tab:compfun} as component functions ($L=6$) for aggregated data generation. As in Simulation 2, the local features of the functions are not so clear by visualization of the generated samples, as it is shown in Figure \ref{fig:sim3} (a). Further, the rules should detect, in addition to the local features of the component functions of Simulation 2, the discontinuity point of Heavisine function and the smoothness of SpaHet function. Figure \ref{fig:sim3} (b) also present the empirical wavelet coefficients of ten generated samples. High empirical coefficients values mean important locations of component functions features to be recovered by the shrinkage rules.

Tables \ref{tab:amse31} and \ref{tab:amse32} show the AMSEs and the standard deviations of MSEs of the considered methods for all the scenarios. As occurred in Simulation studies 1 and 2, the shrinkage rule under logistic prior and ABE had good overall performances for SNR = 3 scenarios. Under that context, the logistic shrinkage rule was the best for Bumps and Blocks functions, while ABE was the best one for Doppler function. The novelty is that BAMS method was the best for Heavisine, Logit and SpaHet functions, although it was not good for the other functions. Since the chosen method estimates simultaneously the six functions, logistic shrinkage rule and ABE would be preferable under this context again. 

For scenarios under SNR = 9, LPM had an overall good performance, being the best one for Bumps, Blocks and Doppler functions, while ABE was the best for estimating the remaining ones. Figures \ref{fig:sim3} (c)-(h) present the boxplots of MSEs for the six component functions in the scenario of $M = 1024$ and SNR = 3. It should be noted that the shrinkage rule under beta prior did not again have good performances and that LPM rule was also not well succeeded in estimating Doppler, Heavisine and the smooth functions Logit and SpaHet.  

\begin{table}[H]
\scalefont{0.7}
\centering
\label{my-label}
\begin{tabular}{|c|c|c|c||c|c|c|c|}
\hline
 \multicolumn{6}{|c|}{\textbf{Simulation study 3} - $\boldsymbol{L=6}$} \\ \hline
  \multicolumn{2}{|c|}{} & \multicolumn{2}{|c||}{$\boldsymbol{n=512}$} & \multicolumn{2}{|c|}{$\boldsymbol{n=1024}$}\\ \hline
\textbf{Function}  & \textbf{Method} & \textbf{SNR = 3} & \textbf{SNR = 9} & \textbf{SNR = 3} & \textbf{SNR = 9} \\ \hline \hline

Bumps &LOG	&	\textbf{0.3899	(0.0262)}	&	0.4188	(0.0118)	&	\textbf{0.2447	(0.0134)}	&	0.2455	(0.0059)\\
&BETA	&	0.4998	(0.0366)	&	0.5840	(0.0206)	&	0.3457	(0.0243)	&	0.4056	(0.0181)\\
&LPM	&	0.4411	(0.0286)	&	\textbf{0.0489	(0.0028)}	&	0.3020	(0.0141)	&	\textbf{0.0334	(0.001)}\\
&ABE	&	0.4032	(0.0268)	&	0.3690	(0.0101)	&	0.2512	(0.0135)	&	0.2191	(0.0045)\\
&BAMS	&	1.374	(0.0609)	&	1.512	(0.0204)	&	0.8199	(0.0255)	&	0.9190	(0.0088)\\ \hline

Blocks &LOG	&	\textbf{0.5138	(0.0402)}	&	0.4015	(0.0151)	&	\textbf{0.4880	(0.0272)}	&	0.2285	(0.0104)\\
&BETA	&	1.176	(0.1184)	&	1.176	(0.0960)	&	1.164	(0.1200)	&	1.007	(0.1235)\\
&LPM	&	0.6998	(0.0451)	&	\textbf{0.0771	(0.0042)}	&	0.8401	(0.0366)	&	\textbf{0.0927	(0.0037)}\\
&ABE	&	0.5243	(0.0407)	&	0.3787	(0.0134)	&	0.4906	(0.0262)	&	0.2173	(0.0089)\\
&BAMS	&	1.382	(0.0793)	&	1.493	(0.0272)	&	0.8439	(0.0411)	&	0.8255	(0.0163)\\ \hline

Doppler &LOG	&	0.8734	(0.0703)	&	0.5543	(0.0310)	&	1.214	(0.0785)	&	0.4121	(0.0207)\\
&BETA	&	1.274	(0.1104)	&	1.071	(0.0727)	&	1.495	(0.1384)	&	0.9149	(0.0890)\\
&LPM	&	1.384	(0.0819)	&	\textbf{0.1543	(0.0100)}	&	2.671	(0.1278)	&	\textbf{0.2944	(0.0119)}\\
&ABE	&	\textbf{0.8518	(0.0637)}	&	0.4450	(0.0240)	&	\textbf{1.213	(0.0759)}	&	0.3647	(0.0183)\\
&BAMS	&	1.534	(0.1068)	&	1.556	(0.0394)	&	1.264	(0.0893)	&	1.028	(0.0318)\\ \hline

Heavisine & LOG	&	5.090	(0.4850)	&	1.164	(0.1098)	&	1.304	(0.0816)	&	0.3282	(0.0215)\\
&BETA	&	5.566	(0.5527)	&	1.889	(0.2260)	&	1.462	(0.1226)	&	0.6707	(0.0736)\\
&LPM	&	11.156	(0.7521)	&	1.226	(0.0756)	&	3.469	(0.1575)	&	0.3852	(0.0180)\\
&ABE	&	4.965	(0.4490)	&	\textbf{0.9368	(0.0810)}	&	1.320	(0.0732)	&	\textbf{0.2709	(0.0188)}\\
&BAMS	&	\textbf{4.296	(0.4739)}	&	2.184	(0.1267)	&	\textbf{0.9275	(0.0877)}	&	0.5113	(0.0256)\\ \hline

\end{tabular}
\caption{AMSE (standard deviation) of simulation study 3 for aggregated data generated with the component functions Bumps, Blocks, Doppler and Heavisine.}\label{tab:amse31}
\end{table}

\begin{table}[H]
\scalefont{0.7}
\centering
\label{my-label}
\begin{tabular}{|c|c|c|c||c|c|c|c|}
\hline
 \multicolumn{6}{|c|}{\textbf{Simulation study 3} - $\boldsymbol{L=6}$ \textbf{(continuation)}} \\ \hline
  \multicolumn{2}{|c|}{} & \multicolumn{2}{|c||}{$\boldsymbol{n=512}$} & \multicolumn{2}{|c|}{$\boldsymbol{n=1024}$}\\ \hline
\textbf{Function}  & \textbf{Method} & \textbf{SNR = 3} & \textbf{SNR = 9} & \textbf{SNR = 3} & \textbf{SNR = 9} \\ \hline \hline

Logit & LOG	&	10.561	(0.8675)	&	2.606	(0.2831)	&	3.794	(0.2898)	&	0.9159	(0.075)\\
&BETA	&	10.228	(0.8989)	&	2.611	(0.3215)	&	3.709	(0.3132)	&	1.186	(0.1754)\\
&LPM	&	23.801	(1.177)	&	2.658	(0.1600)	&	11.931	(0.4723)	&	1.318	(0.0552)\\
&ABE	&	9.756	(0.7668)	&	\textbf{1.659	(0.1897)}	&	3.806	(0.2565)	&	\textbf{0.6812	(0.0562)}\\
&BAMS	&	\textbf{6.454	(0.7391)}	&	2.084	(0.2118)	&	\textbf{2.775	(0.2471)}	&	1.326	(0.0789)\\ \hline

Spahet &LOG	&	3.580	(0.3248)	&	0.8431	(0.1008)	&	7.916	(0.4996)	&	1.851	(0.1427)\\
&BETA	&	3.635	(0.3418)	&	1.021	(0.1392)	&	7.817	(0.5727)	&	2.677	(0.2301)\\
&LPM	&	7.797	(0.5035)	&	0.8733	(0.0526)	&	25.654	(1.059)	&	2.838	(0.1359)\\
&ABE	&	3.398	(0.3027)	&	\textbf{0.6180	(0.0668)}	&	7.934	(0.4271)	&	\textbf{1.100	(0.0975)}\\
&BAMS	&	\textbf{3.319	(0.4242)}	&	1.783	(0.0999)	&	\textbf{4.749	(0.4070)}	&	1.933	(0.1313)\\ \hline

\end{tabular}
\caption{AMSE (standard deviation) of simulation study 3 (continuation) for aggregated data generated with the component functions Logit and Spahet.}\label{tab:amse32}
\end{table} 

\section{Final considerations}
The present paper had the goal to compare the performance of bayesian wavelet shrinkage methods in estimating component curves by aggregated functional data. Simulation studies involving the so called Donoho-Johnstone test functions, the Logit and SpaHet functions as component functions were conducted, for different sample sizes (M = 512 and 1024) and signal-to-noise ratios (SNR = 3 and 9) and number of component curves (L = 2, 4 and 6). 

For scenarios with SNR = 9, the LPM method by Cutillo et al. (2008) had the best performances in terms of averaged mean squared error measure. This result suggests the use of LPM for aggregated functional data with low noise presence. On the other hand, for scenarios with SNR = 3 (high noise presence), the shrinkage rule under logistic prior by Sousa (2020) and ABE method by Figueiredo and Nowack (2001) had good performances. As expected, the overall performances were improved by sample size M = 1024 related to M = 512 but there was not detected meaningful difference in the methods performances for different number of component curves.  

Comparisons with standard and no bayesian shrinkage or thresholding rules, the performances in terms of other measures such as averaged median absolute error and the impact of the chosen wavelet basis are topics of interest for future works.

\end{document}